\documentclass[a4paper]{IEEEtran}

\usepackage[latin1]{inputenc}
\usepackage[T1]{fontenc}
\usepackage[english]{babel}
\usepackage[pdftex]{graphicx}
\usepackage{epstopdf}
\usepackage{amsmath}
\usepackage{amssymb,amsfonts,textcomp}
\usepackage{color}
\usepackage{array}
\usepackage{supertabular}
\usepackage{hhline}
\usepackage{subcaption}

\usepackage{todo}

\usepackage{pifont} 
\newcommand{\cmark}{\ding{51}}%
\newcommand{\xmark}{\ding{55}}%

\newcommand{\ttt}{\texttt}
\newcommand{\North}{\texttt{North}}
\newcommand{\South}{\texttt{South}}
\newcommand{\East}{\texttt{East}}
\newcommand{\West}{\texttt{West}}
\newcommand{\Swallow}{\textit{Swallow}}

\newcolumntype{L}[1]{>{\raggedright\let\newline\\\arraybackslash\hspace{0pt}}m{#1}}
\newcolumntype{C}[1]{>{\centering\let\newline\\\arraybackslash\hspace{0pt}}m{#1}}
\newcolumntype{R}[1]{>{\raggedleft\let\newline\\\arraybackslash\hspace{0pt}}m{#1}}

\title{Overview of Swallow ---  A Scalable 480-core System for Investigating
the Performance and Energy Efficiency of Many-core Applications and Operating Systems}
\author{Simon J. Hollis and Steve Kerrison\\
Department of Computer Science, University of Bristol, UK\\
simon@cs.bris.ac.uk, steve.kerrison@bristol.ac.uk}
\begin{document}

\maketitle

\begin{abstract}
We present \Swallow, a scalable many-core architecture,
with a current configuration of 480 $\times$ 32-bit processors.

\Swallow\ is an open-source architecture,
 designed from the ground up to deliver
scalable increases in \emph{usable} computational power
to allow experimentation with many-core applications and
the operating systems that support them.

Scalability is enabled by the creation of a tile-able system with
a low-latency interconnect, featuring an attractive
communication-to-computation ratio and the use
of a distributed memory configuration.

We analyse the energy and computational and communication
 performances of \Swallow.
The system provides 240GIPS with each core consuming
71--193mW, dependent on workload. Power consumption per instruction
is lower than almost all systems of comparable scale.

We also show how the use of a distributed operating system (nOS) 
allows the easy creation of scalable software to exploit \Swallow's
potential. Finally, we show two use case
studies: modelling neurons and the overlay of shared memory
on a distributed memory system.
	
\end{abstract}

\begin{figure}
\centering
\includegraphics[width=0.49\columnwidth]{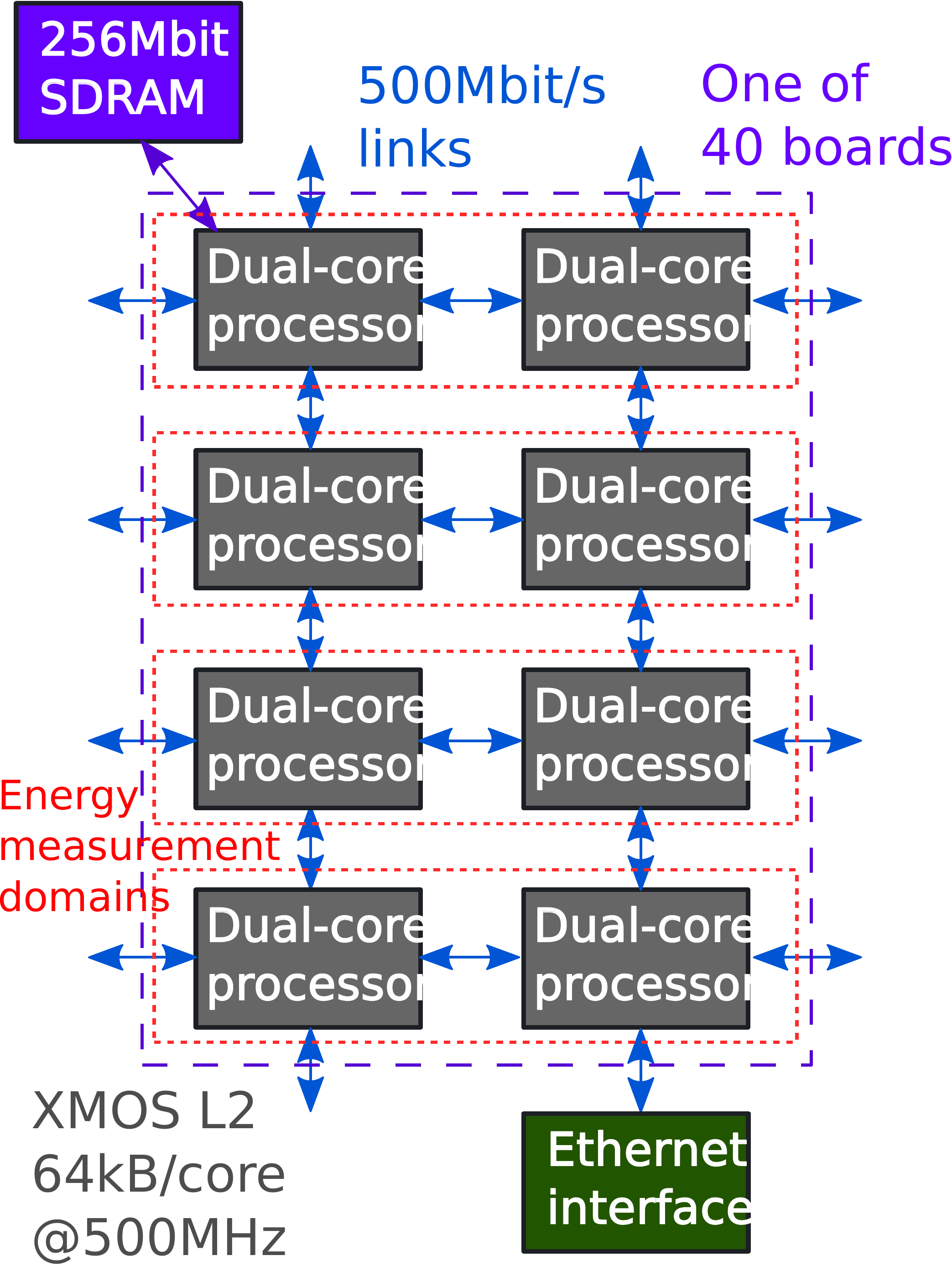}
\includegraphics[width=0.49\columnwidth]{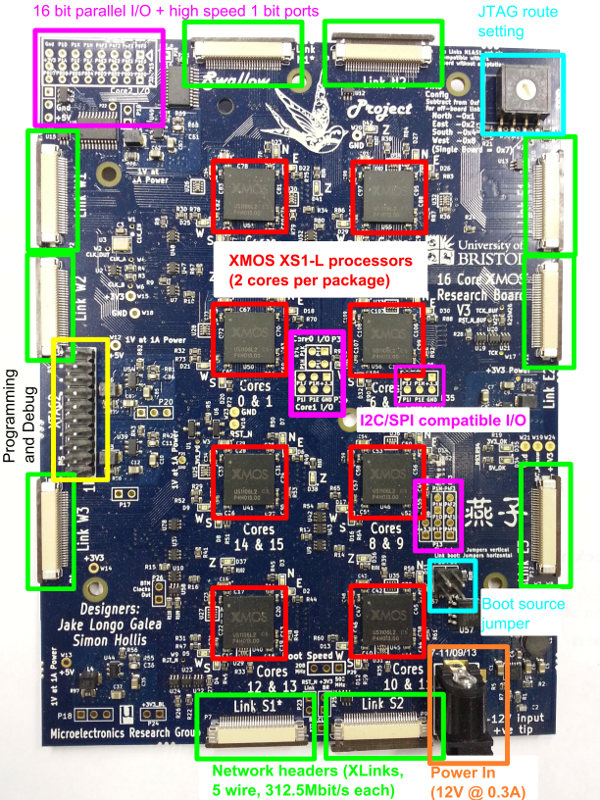}
\caption{A \Swallow\ slice, topology (left); photograph (right)}
\label{dia:Swallow_Slice}
\end{figure}

\section[Introduction]{Introduction}
This paper introduces the \Swallow\ system, a highly scalable many-core system,
which has been developed at the University of Bristol, UK.

\Swallow\ has been developed as an experimental system for investigating techniques
to progress energy efficient and scalable parallel systems.
Key aims in the use of \Swallow\ are:
\begin{enumerate}
\item To experiment with new techniques for parallel programming
across hundreds of processing cores;
\item To factor energy consumption into both user programming
paradigms and operating system decisions;
\item To develop a high-performance energy-aware operating system
to support thousands of processing threads across hundreds
of processing cores with minimal user interaction.
\end{enumerate}

To support these aims, we needed to build a scalable and analysable
many-core system, augmented with energy measurement and good
processor-to-processor communication performance. \Swallow\ is the
realisation of this need.

Before we go on to explain the technical details of \Swallow, we consider the need
to make many-core systems and associated programs and operating systems.
Computational needs are growing and many-core has been
readily accepted as an energy- and cost-efficient method of continuing this
growth, especially as we reach the limits of Moore Scaling. It is also well known that only some applications will parallelise to the extent needed to exploit a large-scale
many-core system. 

In the design of \Swallow, we assumed two use cases: first, the conventional case
where a user develops a highly parallelised implementation of a single
application. Second, and since to exploit all 480 cores of Swallow, most such
tasks would need to be in the ``embarrassingly parallel'' category~\cite{Bailey:1994}, we also support
multiple non-interacting applications running simultaneously. This allows more
efficient system utilisation.

During the development of \Swallow, we used a number of applications
as motivating examples to evaluate the impact of our design decisions.
Some are highly parallel, such as image processing, neural networks and streaming matching
algorithms. Our other goal of many simultaneous tasks was supported
by the development of a distributed
multi-tasking operating system, nOS, which supports a number of smaller
processes. nOS is detailed in another paper~\cite{Hollis:nOS}, so we
focus on the category of parallelised algorithms in this paper.

\section[Requirements for Scalable Many-core]{\rmfamily Requirements for
Scalable Many-core}

To ensure that the features of the architecture used for evaluation of
programming do not interfere with the fundemental properties that we 
are trying to explore, it is necessary to build a scalable architecture.

Building scalable many-core systems is an area well focused upon by the academic
community, yet large-scale real and usable implementations are few and far
between. The reason behind this is the fact that to build a scalable many-core
system one needs: 

\begin{itemize}
\item to source or create many components, such as processors, networks and
memory systems that differ from those commonly available for single or
multi-core systems; 
\item fresh programming practices; and
\item a large investment of time and engineering effort.
\end{itemize}

We have solved the above problems in the following way:

\begin{itemize}
\item the use of a novel commercial processor architecture, which integrates on-
and off-chip networks as a first-class primitive, in combination with a
simplified memory hierarchy;
\item the use and extension of a recently developed multi-core commercial
programming language and the creation of design patterns for efficiently
exploiting the underlying hardware's capabilities; and
\item a large investment of time and engineering effort.
\end{itemize}

\subsection[Scale-free systems]{Scale-free systems}
\label{sect:Scale-free}
Work extending from the PRAM model of computation (e.g. Valient, Gibbons, Culler~\cite{Valient:1990,Gibbons:1989,Culler:1996}) shows a good agreement in the
academic community that a \emph{scale-free} computing system must have the
following five properties:

\begin{enumerate}
\item Independent processors i.e. the behaviour and timing of any processor
should not depend on the current activity of any of the other processors.
\item Storage where the total capacity per processor remains at least constant
under scaling.
\item Storage where the access time is independent of the number of processors
\item A communication system whose capacity scales at least linearly with the
number of processors and with a constant or predictable latency.
\item Predictable execution timing if processor-to-processor synchronisation is
needed~\cite{Valient:1990}
\end{enumerate}

The establishment of the above properties yields a system where the performance
of a single processor in the system is not constrained by the actions of other
processors. Thus, each processor in the system can operate up to its maximum
capacity and a system of $N$ processors, each with processing capacity
$c_i$ can therefore have a maximum theoretical processing capacity of
$\sum^{N}_{i = 1} c_i$. Note that this formulation is
independent of whether or not the system is homogeneous or heterogeneous.

\subsection{Scalable communication}

Work by Valient, and extended by May~\cite{Valient:1990,
May2008} shows that for each application in a parallel system there is
a limiting bound on performance. Given a node's data source and sink throughput
$e$ and that node's communication throughput $c$, and overall system data
source and sink throughput $E$ and total communication throughput $C$, this
can be expressed as:
\begin{equation}
\mathrm{Communication\ peformance} =  \max\left(\frac{e}{c},\frac{E}{C}\right)
\label{eqn:e_over_c}
\end{equation}

Therefore, we can formally define the amount of communication bandwidth required to ensure
that communication does not throttle performance as:
\begin{equation}
\frac{e}{c}\ \le \  1 \ \ \ \wedge \ \ \  \frac{E}{C} \ \le\  1
\label{eqn:e_over_c_gt1}
\end{equation}

However, it is not always necessary to build a system that gurantees both of these properties.
Rent's Rule~\cite{Landman1971} and the Berkeley Dwarves~\cite{Asanovic:LandscapeofParallel} give
both a theoretical and empirical insight into the fact that the majority of data transfers
in a parallel system are localised. Therefore, whilst we likely still desire a system
with $\frac{e}{c} \le 1$ in most circumstances, the system long range capacity may not need to be as large as suggested by Eqn.~\ref{eqn:e_over_c_gt1}; nor need it have uniform capacity for most applications.

These observations also throw into question the usefulness of assertions, such as those
made by the advocates of the PRAM model~\cite{Valient:1990, Gibbons:1989}:
that a scalable system must have at least a $log_p$
scaling of communication latency with system size --- whilst this is clearly necessary for
systems with global communication or synchronisation, systems with only local traffic
do not require such aggressive network performance and, for most real application scenarios
with limited farmer-worker, or pipelined execution patterns, high performance
local interconnect will provide a close match to the requirements.

For this reason, \Swallow\ has been initially evaluated with a 2D mesh network structure.
\Swallow\ supports richer 2D and 3D networks, but at this stage we judge it unnecessary
to implement these given our choice of application domains: our target of
supporting multiple disjoint applications implies a degree of locality in communication and
farmer-worker computations also demonstrate locality.

\subsection[Scalable memory]{Scalable memory}
\label{sect:Scalable_Memory}
The main bottleneck in modern computing systems is the memory sub-system.
Compared to processing elements, most memories have very high latency and
limited data bandwidth. This makes them the most important components to
design for a scalable architecture.

For ease of programming, \emph{shared memory} architectures are very common,
in which multiple processors are interconnected to a shared storage structure.
The demands on this interconnection are very high --- commonly it must be a
full cross-bar, limiting scalability. Further, contention on the memory
system is increased and this reduces its average case performance for
both latency and per-processor bandwidth.

There have been many optimisations to this problem, such as adding levels of
independent caching between processors and the shared memory, but the shared
memory model still demands that there be consistency in the view of each
processor to the values stored in the shared structure. This has led to the
development of cache coherency protocols, about which there is a wealth of
literature. However, whilst advances have been made, the problem of coherent
shared memory is fundamentally unscalable in the long term. This is evidenced by
the very limited number of processors that shared memory systems tend to have
(up to about 16).

Clearly, this will not meet our goal of scaling to thousands of processors,
so we need a different approach, which we outline in Sect.~\ref{sect:Memory}.

\subsection{Summary of scalability requirements}
The requirements outlined in \S~\ref{sect:Scale-free} give rise to the following design decisions in Swallow:

\begin{itemize}
\item Non-interacting instruction execution. Each processor performs computation
independently, with any external interactions being explicitly defined in the
programming language.
\item Processors do not share memory, rather each processor has its own,
independent memory. This approach ensures that a processor's execution cannot
depend on another processor's memory access pattern.
\item A simple memory hierarchy. Keeping the memory hierarchy simple ensures
that memory accesses remain predictable.
\item Explicit communication of data values between processors. Making
communication explicit allows clear analysis of required data rates and the
impact of communication latency on a program. It also allows analysis of data
paths and aids allocation of communication resources.
\item The communication network is well-provisioned in comparison to the volume
of data that can be injected by the processors present in the system. Having a
well-provisioned network reduces the chance that the network itself becomes a
bottleneck and that data from one processor's communication can interfere with
that from other. Therefore processor independence is maintained.
\item The communication network can be set to run in circuit-switched mode,
which offers predictable and non-blocking communication.
\end{itemize}

The net effect of our design decisions is to create a system that is not only
scalable in terms of processors, memory and network capacity, but also
facilitates predictable execution, with the gains in system efficiency and
static analysis that result.

In the following sections, we outline the details of the implementation that
follows these decisions, and show how they result in the gains asserted above.

\section[Memory subsystem]{Memory subsystem}
\label{sect:Memory}
In designing \Swallow, we had a very specific goal in mind: to create a system
that does not re-create any of the scalability problems suffered by most
commercial and experimental many-core systems (see \S~\ref{sect:Scalable_Memory}). It has been demonstrated time
and time again that creating efficient shared memory is the most difficult
problem that these systems present. Shared memory causes bottlenecks when
consistency must be preserved across even small numbers of processing nodes,
and with Swallow we wished to build a system that scales to many thousands.

This necessitates aggressive design decisions and, at an early stage, the design
decision was made to build Swallow as a distributed memory system. Each core
possesses its own private memory. Amongst other things, this gives us a
guarantee of non-interference. In order to ensure scalability of software, as
well as hardware, we wished to extend this to provide predictability of access
times too, and for this reason, Swallow is also a cache-free design that has a
single level of memory hierarchy above the registers.

Taken together, this ensures that every processing node in Swallow has
predictable memory and computational timings, easing the job of a programmer in
creating balanced and scalable distributed computations. An effect of this
decision is that processing nodes have direct access to only 64kB of store,
which is shared between instructions and data.

Swallow also incorporates a single 256Mbit DRAM on each slice, which is attached
to one of the cores on the board as an I/O device. The DRAM can then be
accessed by running a thread on this core with a memory controller
functionality. We have developed code that is able to emulate the hardware
functionality of a larger memory in software. A memory coordinator process listens for memory
address requests, coming as messages from Swallow cores and either commits or
returns data values in response to this. Note that, from the point of view of
an accessing core, this makes the DRAM appear in just the same way as any other
processor; the task that it is running is to load and store data.
Therefore, the assumptions about predictability of individual cores' timings
and local memory accesses still hold.

\begin{figure}
\centering
\begin{minipage}[b]{0.55\linewidth}
\includegraphics[width=\columnwidth]{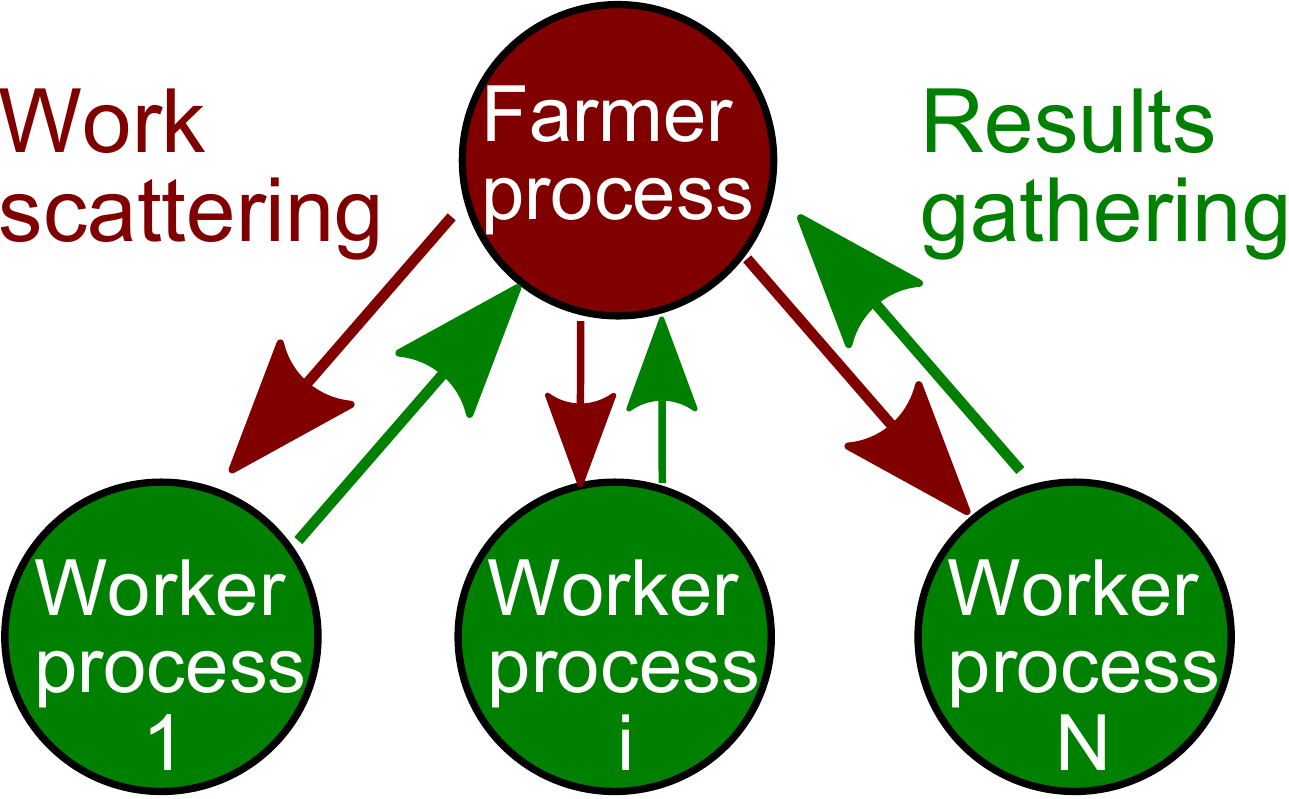}
\vspace{0.5em}
\subcaption{Farmer-worker computation}
\label{dia:farmer-worker}
\end{minipage}
\hspace{0.5em}
\begin{minipage}[b]{0.24\linewidth}
\includegraphics[width=\columnwidth,clip,trim=1cm 11.5cm 12cm 0.5cm]{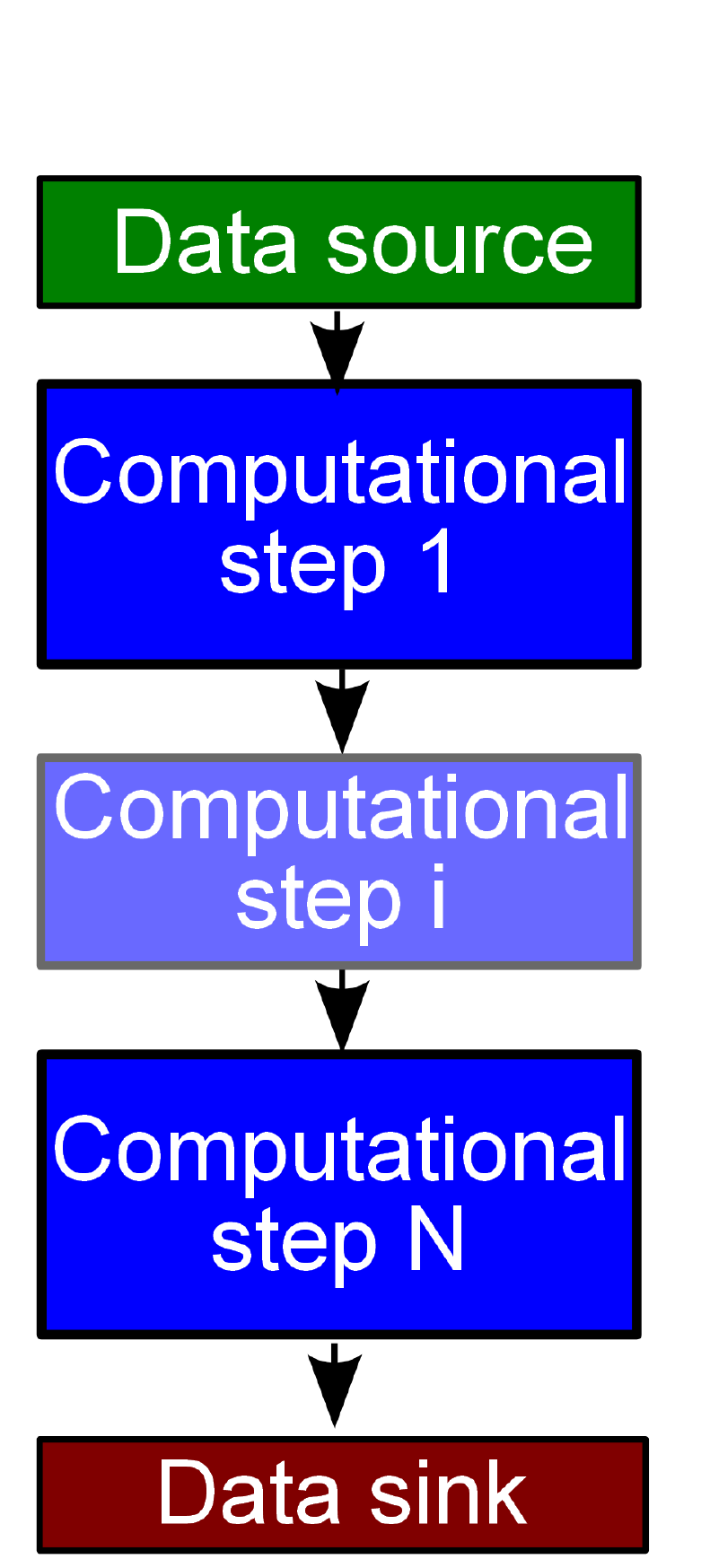}
\subcaption{Pipelined}
\label{dia:Pipelined_Computation}
\end{minipage}
\caption{\Swallow\ target computational paradigms}
\end{figure}

Whilst the size of local stores would appear a limiting factor for both programs and data, this is not often the case in practice. The program size is not a fundamental limiting factor, since any computation can be structured as a series of computations, spread across multiple cores, producing `pipelined' computation (see Fig.~\ref{dia:Pipelined_Computation}). 
Data storage capacity is also not a fundamental limit on the applications that may be
supported. Large data sets can be addressed in a number of ways:

\begin{enumerate}
\item A \textit{farmer-worker} (or \textit{scatter-gather}) approach can be used to split the data into
sub-sets and each sub-set assigned to a separate processor, each executing the
same program. This also requires a coordinating node and appropriate I/O (See
Fig.~\ref{dia:farmer-worker}).
\item The computation can be arranged as a set of \emph{streaming
computations}, where only a small part of a larger data set is ever stored at a
processing node at one time. A much larger proportion of the set is stored
across all nodes involved in the computation (blue line, Fig.~\ref{dia:Memory_Scaling}),
and the entirety of the data set is then stored either
externally (via Swallow's network interface) or on Swallow's on-board, but
off-processor DRAM.
\end{enumerate}

We assert that streaming is much preferred from an efficiency
point of view, since it introduces no
centralised point of contention. However, many contemporary algorithms and
high-level programming languages are constructed using the scatter-gather approach,
so we must support this too.

\subsection{Nodes as remote data storage}

We also make a key observation in the construction of a large-scale distributed
memory system: a node where no processing takes place can be used instead as
a remote data store. Thus, \textbf{any data size can be stored at the expense
of a proportional number of processors}. In a small-scale system, sacrificing
processing ability is unlikely to deliver a satisfactory throughput. However,
in a system like Swallow with hundreds of nodes, it is likely that some
processors will be unusable for a given application, and their resources can be
dedicated to storage with minimal impact on the underlying computation. i.e. we
can support many, small computations or a few large ones \emph{without changing the underlying processing fabric}. Our approach of
deploying idle cores as memories also presents a solution to the
dark silicon problem where, in some many-core systems, their is 
insufficient power or thermal budget to fully utilise all processors
at once.

\begin{figure}
\includegraphics[width=\columnwidth]{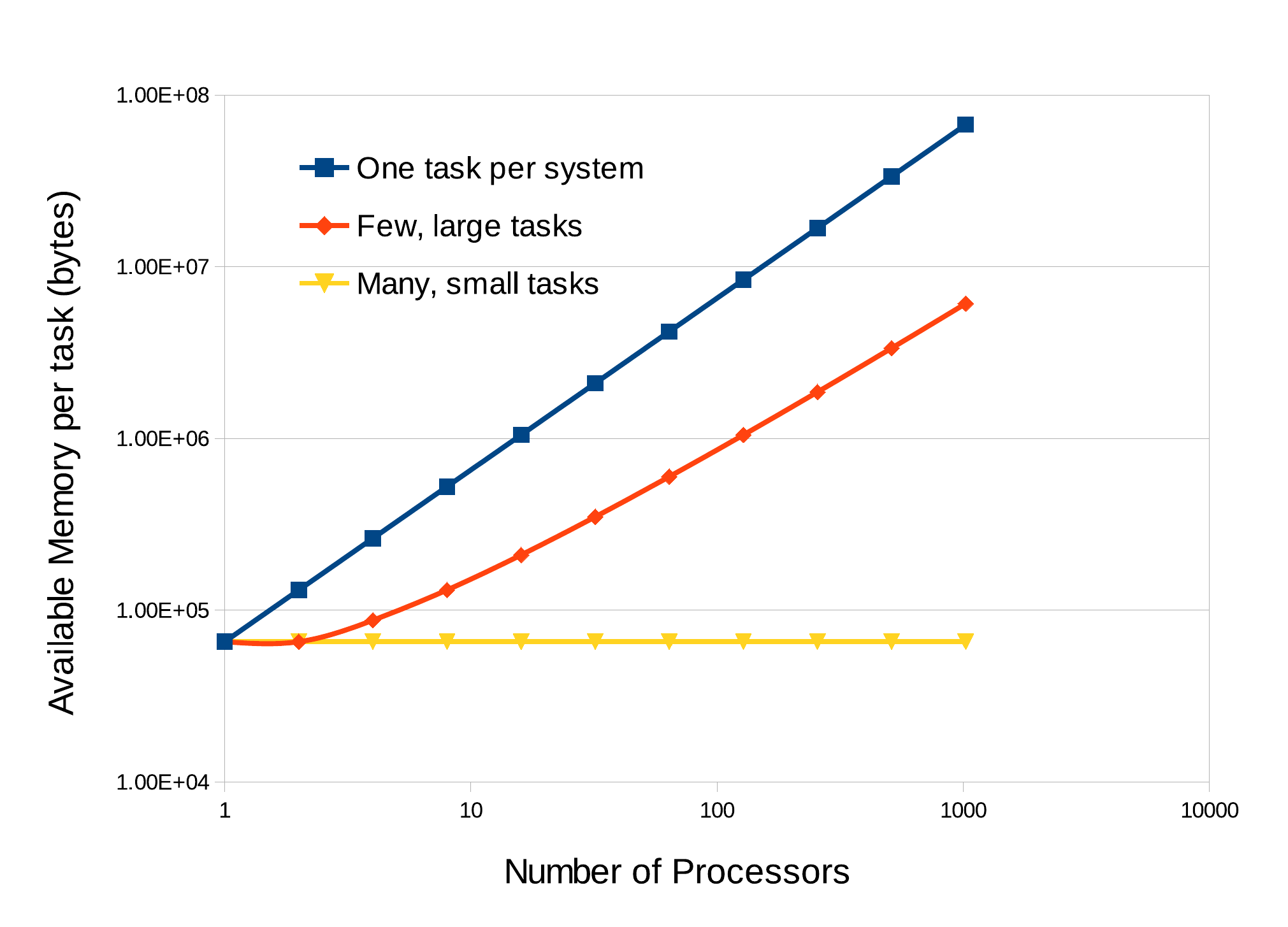}
\caption{Memory available per task with number of tasks.}
\label{dia:Memory_Scaling}
\vspace{1mm}
\end{figure}

In Fig.~\ref{dia:Memory_Scaling}, we show how Swallow can
support multiple use cases when balancing computations and memory. If we
consider building Swallow systems of exponentially increasing size, we see that
we can either build a system with a single tasks and an exponentially
increasing amount of memory (blue line), or an exponential number of tasks, each with a fixed memory size of 64kB (yellow line).

A third example (red line) shows what happens if the number of tasks scales
linearly with exponential growth in the number of processors; there is a slower-than-exponential growth in memory per task, yet more tasks are supported. These three curves illustrate that a full continuum of process
number and memory requirements can be supported on \Swallow.


\subsection{Remote code storage}

\begin{figure}
\includegraphics[width=\columnwidth, trim=150 300 150 40]{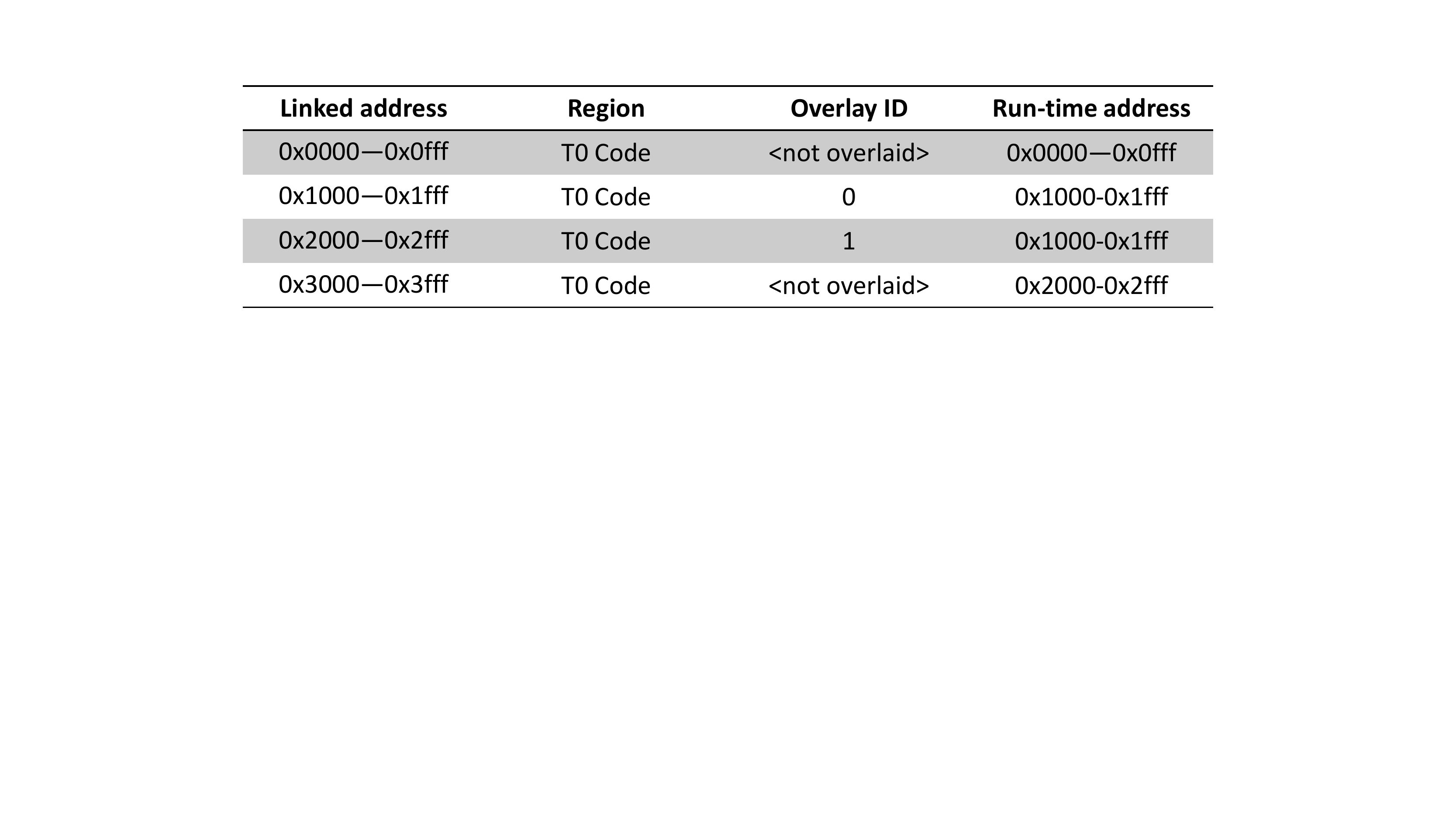}
\caption{Code overlays and how they relate to memory addresses}
\label{dia:Code_Overlays}
\end{figure}

With additional software handlers, it is also possible to use the larger data
storage of the DRAM to support larger code sections than can fit into local memory
via the use of code \textit{overlays}. Overlays allow
run-time swapping in and out of code segments on the individual processors via
a series of interrupt handlers that trigger when execution crosses an overlay
boundary (Fig.~\ref{dia:Code_Overlays} shows one potential mapping for a system with 4k word
overlays). The program region is 16k-words in size, but overlays allows the range between \ttt{0x1000--0x2fff} to be overlaid into $2\times4$k-words, reducing the overall program size
to 12k-words. Note the similarity between overlays and
a virtual memory system with on-demand paging. The difference with overlays is that
only some parts of the program are subject to overlaying and accesses
are still based on physical addresses. This means timing
critical sections of code can be guaranteed not to be changed by simply disallowing them
to be overlaid.

However, we do not recommend the use of overlays: the use of interrupts and the need to wait
during execution for additional instructions to be loaded across the network
both result in a reduction of predictability in the target application. It is
better to re-engineer the application to make use of either fewer instructions, less static data or more processors. If this cannot be done,
then overlays provide an alternative solution.

\section[Design selection]{Swallow overview}
In this section, we explain the design choices made when realising \Swallow.

\subsection[Scalable construction]{Scalable construction}

For economic and reliability reasons, we decided to construct \Swallow\ from
\textit{slices}. A slice is shown in
Fig.~\ref{dia:Swallow_Slice}. Each \Swallow\ slice comprises
sixteen processors, each with their own memory, 10 bi-directional on-board
network links and 10 off-board network links.

We highlight the PEs (red),
SDRAM and SPI interfaces (magenta), slice-to-slice links (green) and debug interfaces (yellow and blue). The card measures 105mm wide x 140mm high, and consumes a maximum of 5W at 12V. Multiple slices can be stacked in 3-D or spread out in 2-D, using the mounting holes and flexible FFC-type cables. This allows physical rearrangement of both the boards and the links. Using this
arrangement, both 2-D and 3-D network topologies can be created and grid and
non-grid topologies may be used.

We built 40 slices of \Swallow, enabling the construction of a 640 processor
system, however yield issues (mostly with connectors) mean that the largest
machine we have been able to build and fully test is 480 cores.

\subsection[Processor choice]{Processor choice}
A design decision was made to use commercially available parts wherever possible
in the construction of \Swallow. For processor selection, this meant evaluating
the commercially-available cores. We searched for systems with the scalable
computational properties outlined in S\ref{sect:Scale-free}. 
According to our definition, this requires a
data path with timing predictability, simple flat memory and a rich interconnect.
A summary of our findings is in Tab.~\ref{tab:core_selection}.

\begin{table*}
    \centering
    \begin{tabular}{|l|L{1.35cm}|L{1.6cm}|l|L{2.30cm}|L{2cm}|L{1.6cm}|C{1.25cm}|}
        \hline
    \textbf{Processor} & \textbf{Cores $\times$ data width} & \textbf{In-order
    pipeline type} & \textbf{Cache} & \textbf{Memory configuration}  &
    \textbf{Multi-core interconnect} & \textbf{Time deterministic} &
    \textbf{Overall suitability} \\
\hline
ARM Cortex M & $1\times$32-bit  & Single-scalar & Optional & <varies>  & No~\xmark & If no caches~\cmark & \xmark \\
\hline
ARM Cortex A, single core & $1\times$32-bit  & Super-scalar & Yes & <varies> & No~\xmark & No~\xmark & \xmark \\
\hline
ARM Cortex A, multi-core &  $4\times$32-bit  & Super-scalar & Yes & <varies>  & Memory coherency~\cmark & No~\xmark & \xmark \\
\hline
Parallela Epiphany & $64\times$32-bit & Super-scalar  &No & Local + global SRAM
& Network-on-Chip~\cmark  & No~\xmark & \xmark \\
\hline
XMOS XS-1 & $1\times$32-bit  & Single-scalar & No & Unified, single cycle SRAM & Network-on-Chip~\cmark & Yes~\cmark & \cmark \\
\hline
%
MSP430 & $1\times$16-bit   & Single-scalar  & No  & I-Flash + D-SRAM  & No~\xmark  & Yes~\cmark  & \xmark \\
\hline
AVR & $1\times$8-bit   & Single-scalar  & No  & I-Flash + D-SRAM  & No~\xmark  & No~\xmark & \xmark~\\
\hline
Quark & $1\times$32-bit  & Single-scalar  & Yes  & Unified DRAM  & Ethernet~\xmark  & No~\xmark & \xmark \\
\hline
\end{tabular}
\caption{Comparison of candidate \Swallow~\ processors (suitable: \cmark; unsuitable: \xmark). Only the XS-1 meets all requirements}
\label{tab:core_selection}
\end{table*}

As can be seen, few commercial processors present the necessary characteristics
of a scalable architecture.  The XMOS \hbox{XS-1} architecture is the only candidate to provide all of these. We further investigated the architecture and determined that its feature set was an excellent match for our system requirements.

Here are the key technical characteristics of the XMOS \hbox{XS-1} architecture:

\begin{itemize}
\item In-order single-scalar processor
\item Every instruction completes in one cycle (except port- or network-based
I/O, which may block)
\item Overhead-free context switching of up to eight hardware threads
\item Network input and output as ISA-level primitives
\item Memory accesses are single cycle
\item Predictable execution at all levels means that thread timing is statically
analysable
\item A built in network, which extends on- and off-chip.
The network supports up to $2^{16}$ nodes, and its operating speed can be
controlled in software.
\end{itemize}

There are five widely available devices based on the \hbox{XS-1} implementation. Three
single-core, one dual-core and one quad-core device. To maximise the density of
the final configuration, whilst leaving power budgets modest, the dual-core
device was selected (XS1-L2A). The rated operating frequency of this device is
500MHz, so each core offers 500 million integer operations a second, which is
shared across at least four threads. The maximum per-thread rate is
125MIPS.

\subsection{Thread throughput}
\begin{figure}
\centering
\includegraphics[width=\columnwidth]{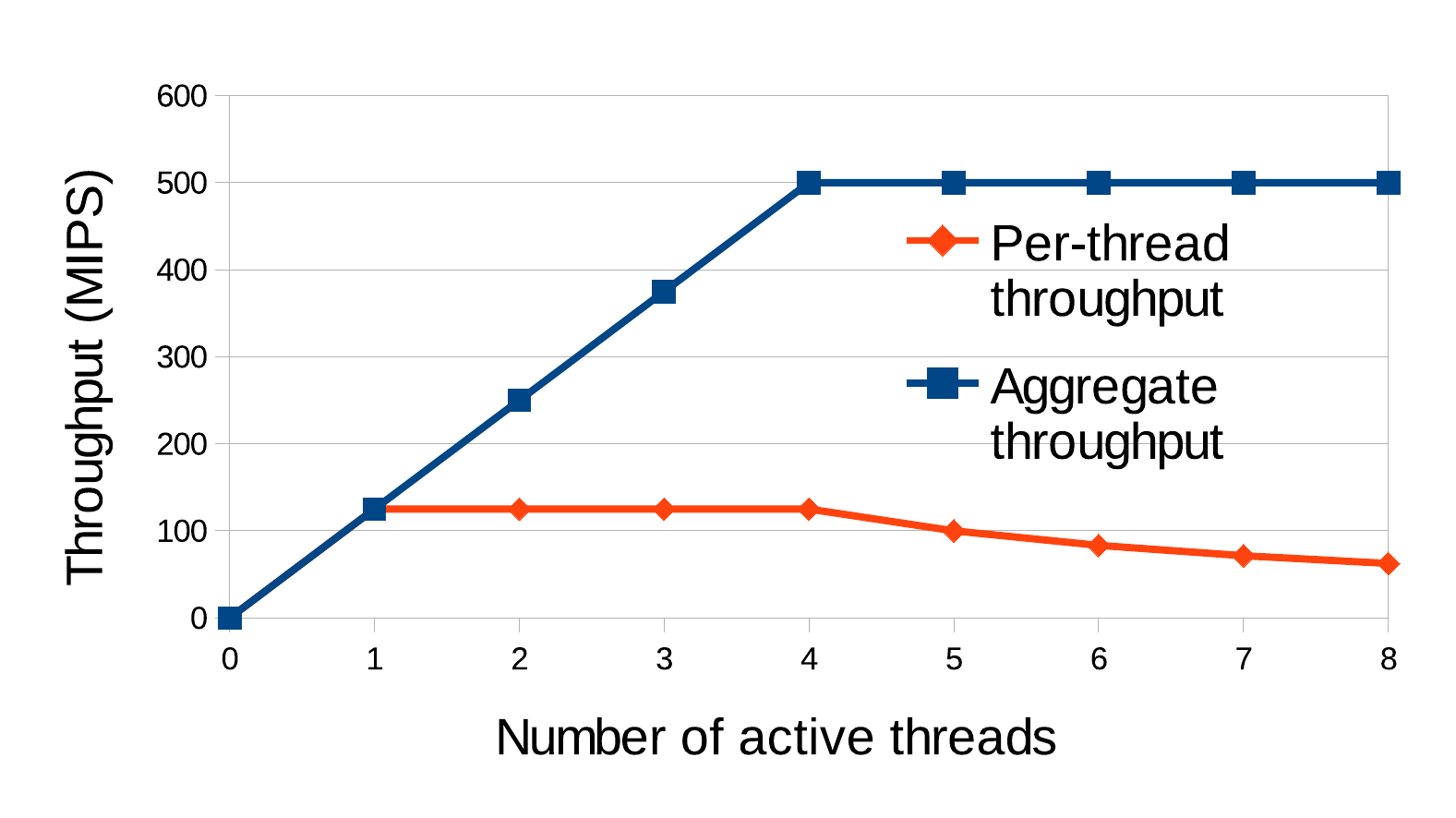}
\caption{Scaling of thread throughput with number of threads on the \hbox{XS-1} architecture (at 500MHz clock speed)}
\label{dia:thread_throughput}
\end{figure}

Fig.~\ref{dia:thread_throughput} shows that the per-thread throughput
and processor aggregate throughput
scales predictably, according to the number of threads active (i.e. in the running state) in the system. The data is taken from~\cite{XMOS2014} and shows the effect of the 4-slot pipeline in the
\hbox{XS-1} architecture: per-thread throughput if constant for up-to four threads, then
declines linearly. We also see that processor aggregate throughput is maximised when
at least four threads are active.

\subsection[Network-on-Chip Implementation]{\rmfamily Network-on-Chip
Implementation}

\Swallow\ builds on the network that is included in the XS-1 architecture. In this
architecture, each processor is supplied with four external communication
links. Links are flexibly allocated, with partitioning and routing being set in
software. Links can be aggregated to form a single logical channel with
increased bandwidth; or connected separately to different parts of the network,
to include the dimensionality of the resulting network topology. They may also
be left unused. Links connect to 12-ported switches, one per core. Routing selection at switches is implemented as a series of
longest-prefix address comparisons and is very efficient, leading to low-latency
routing.

\section{Swallow interconnect}
Swallow contains a rich, high performance interconnect, suitable for supporting
arbitrary traffic types and a mix of parallel or non-parallel applications. In the
following sections we give an overview of the interconnect and implementation details.

Given our three target use cases of farmer-worker, pipelined and multiple independent applications, it is not necessary to produce a fully-connected interconnect topology.
Communication patterns of the use-cases exhibit high spatial locality and
clustering of associated processing elements. In this case, topologies
such as a 2D mesh that are not universally scalable remain useful in providing
acceptable latency and bandwidth in a many-core system.

This has the beneficial side-effect that expensive high-degree topologies (e.g.
Clos, hypercube) become unnecessary to show the fundamental properties under investigation. With \Swallow, this allows us to simplify the topology with
fixed degree, independent of the number of nodes.

\subsection[Network topology]{Network topology}
\begin{figure}
\includegraphics[width=\columnwidth]{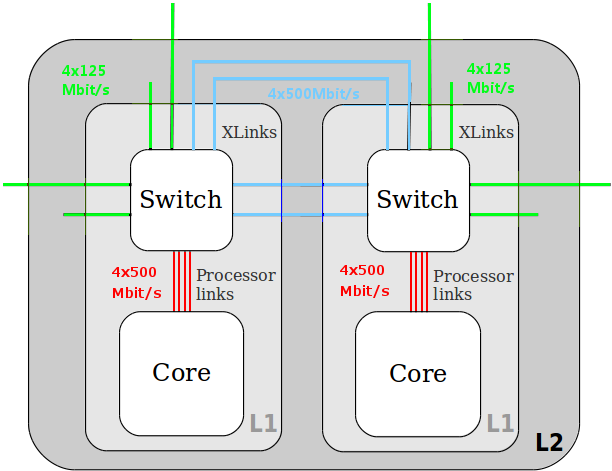}
\caption{Network link configuration for a single \Swallow\ node}
\label{dia:XS1-L2A_links}
\end{figure}

The device chosen for the Swallow system contains two processing cores and
exposes four network links, in the way shown in
Fig.~\ref{dia:XS1-L2A_links}. The two internal links contain
four times the bandwidth of the external links and data words can be
transferred from the core to the network hardware with \textit{just three
cycles} of latency (6ns). 
This compares with 80ns for the BlueGene/Q system~\cite{Kumar2012}. 
In Swallow, the external links are
then arranged to connect \North, \South, \East\ and \West\ to other devices, which are notionally arranged in a
grid pattern.

\begin{figure}
\includegraphics[width=\columnwidth]{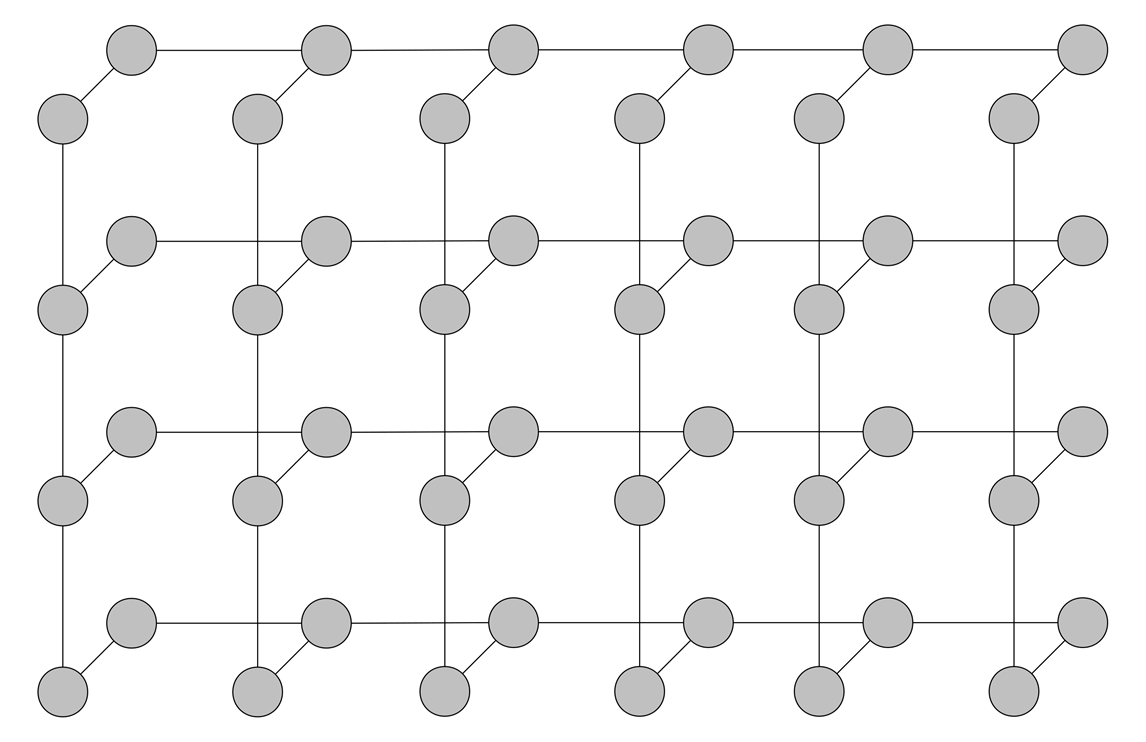}
\caption{\Swallow's `lattice' network topology}
\label{dia:Brick_Wall}
\end{figure}

An interesting artefact of \Swallow's device selection is that, as seen in
Fig.~\ref{dia:XS1-L2A_links}, it is not possible to make a
conventional mesh topology. The internal links already utilised by the
core-to-core connection mean that an attempt to create such a grid in-fact
results in the creation of a `lattice' structure, resembling that shown in
Fig.~\ref{dia:Brick_Wall}. This presents interesting
routing challenges, since a default X-Y approach will not lead to a fully
connected network. To solve this, we implemented a new network routing table
generation tool, which uses dimension-ordered routing to solve the connectivity
problem. 

The network is effectively composed of two layers, with each layer containing
half of the available cores. One layer routes in the vertical dimension and the
other layer routes in the horizontal dimension.  Each node in the network also
has a connection to a node in the opposite layer, which takes place within a
chip package. This topology requires that 2D routes be translated into a form of
2.5D routing, where routing between layers is required to change
horizontal/vertical direction. The dimension order routing strategy that we use
prioritises the vertical dimension first. If a node is attached to the
horizontal layer and a vertical communication is required, the data must
therefore be sent to the other layer first. In this scheme, there will be at
most two layer transitions; the case being two nodes attached to the horizontal
layer that do not share the same vertical index.

Swallow links use flexible cables, allowing the physical topology of the network to be adjusted across a wide variety of configurations, further extending the range of experiments that can be carried out. New routing algorithms are simply programmed
in software to cope with these.

\subsection[Network capacity]{Network implementation}

\Swallow\ nodes contain one processor, one switch and 12 network links.
Each network link in Swallow is physically formatted over five wires in each
direction and uses wormhole routing with credit-based flow control. This is
abstracted at the instruction set level into channel communication which can
take the form of either channel switched or packetized operation.

\textbf{In packet operation}, routes are
opened with a three byte header prefixed to the front of the first token emitted
from a communication channel end. Any network links utilised along the route in
that direction are held open by the channel until a closing control token is
emitted.  If the close token is never emitted, the links are permanently held open,
effectively creating a dedicated channel between two endpoints. This is
the mechanism for realising channel switching.
Packetized transfers incur an overhead from the header and switch setup time, resulting in a typical effective data rate of approximately 435Mbit/s, depending on packet size.

\textbf{Channel switching} increases the effective data rate to 500Mbit/s
by removing header and
control token overheads, but physical links become unavailable to other
channels. Multiple links can be assigned to the same routing direction where a
new communication will use the next unused link. This increases bandwidth,
provided the number of concurrent communications in that direction is equal to
or greater than the number of links.

\subsection{Network details}
Each
link uses the same signalling protocol, but may run at different speeds.
The link connections and speed settings used in \Swallow\ are shown in Fig.~\ref{dia:XS1-L2A_links}.
We can see that internal bandwidths (red and blue) exceed external bandwidths (green)
by a factor of four and allows a node to be configured with an inactive processor and used only for routing, with no impact on the overall bandwidth of the resulting network.
Data has a maximum throughput of 500Mbit/s per internal link and 125Mbit/s per external link,  giving a device package
internal bandwidths $c$ of 2Gbit/s and external bandwidth of 500Mbit/s.


Links send data in eight-bit tokens comprised of two-bit symbols. A token's
transmit time is $3T_s + T_t$, where $T_s$ is the inter-symbol delay and $T_t$
the inter-token delay, measured in switch clock cycles. The fastest possible
mode is $T_s = 2, T_t = 1$, yielding the aforementioned 500Mbit/s at 500~MHz.
The external links must be run with larger delays to preserve signal integrity,
giving 125Mbit/s throughput per external link.

The total core to core latency for an eight-bit token is 270~nS. The total core
to core latency for a 32 bit word between packages is 360~ns, equivalent to the
time taken to for the sending thread to execute 45 instructions.  Between two
cores in a package, this reduces to 40 instructions.  Core-local channel
communication can take place in 50~nS, or approximately 6 instructions. These
timings are taken following a synchronisation between communicating channel
ends, which incurs a small overhead.

\subsection[Ratio of communication to computation]{\rmfamily Ratio of
communication to computation}

%
%
%
%
%
%
%
%
%

We can calculate the theoretical maximum ratio of communication to computation
for \Swallow. It is possible for a single thread of
execution on a \Swallow\ processor to issue 125MIPS. Byte or 32-bit long communication
instructions complete in a single cycle, implying a maximum per-thread
communications throughput 4Gbit/s. There are up to four threads, so the maximum
process demand is 16Gbit/s.

However, in reality the serialisation of the data onto and from the network
requires one cycle per byte at a frequency of 500MHz (External links 125MHz) to traverse the
network interface. Any communication instructions of the same type issued from
any thread within this time will block until the network interface is again
free. Thus, the maximum available rate of issuing communication from the
processor is one byte sent and one byte received every cycle 
(every 2ns [External 8ns cycle]) across all threads.

This information is known at compile time, and thus well-structured programs for
\Swallow\ will take this into account, giving a new calculation. Therefore, we
can calculate that the maximum $e$ value of a well-formed set of threads in \Swallow\
is 4Gbit/s. $c$ is 2Gbit/s, implying an $e/c$
ratio of 2. 

The available data rate on device external network links is one quarter that of
the internal links (a byte requires 64ns to transmit, giving a peak data rate
of 125Mbit/s). In the case of high levels of network traffic, this means that
the network will either congest at these interfaces, if the same route is
required by all four threads, causing $E/C= 32$, or in the case that all
four streams take different paths, the congestion does not occur and $E/C=8$. So, globally, \Swallow\
provides a range of execution to communication throughputs of $8 \le E/C \le 32$.

\subsection[External network interface]{External network interface}
Communication into and out of \Swallow\ is performed by the use of an ethernet
bridge module. This module attaches to the \Swallow\ network and is addressable
as a node in the network, but forwards all data to and from an ethernet
interface. Using this bridge, we are able to bootstrap, load programs and data
into \Swallow\ as well as to stream data in and out of it at run time.

\Swallow\ supports any number of ethernet interfaces, with up to two per slice
being supported (on the \South\ external links). Each bridge can
support up to 80Mbit/s of data transfer in each direction, so this places a
limit on what proportion of the overall network traffic can be external traffic
($\sim \frac{80}{125}$). This limit is only due to the construction of the bridge module software --
future versions could potentially match the total network capacity.

\section[Energy measurement]{Energy measurement}
\Swallow\ was designed to be an energy-transparent system. To this end, a number
of power measurement points have been designed into the system. Swallow cores
are powered by five separate switch-mode power supplies, each fed from a main
5V switched supply. Four of these supply 1V to the cores, with each supply
powering two devices (four cores). The fifth supplies I/O pins across all
devices, as well as the ancillary reset and configuration circuitry. 

Each power supply has a shunt resistor on its output and associated probe
points. We created a daughter-board
that incorporates sensitive differential voltage amplifiers and a high-speed
multi-channel analogue-to-digital converter. The resulting system is able to
measure individual power supply energy consumption at up to 2MSps, or 1MSps if
all supplies wish to be sampled simultaneously. The schematic for the system is available online as part of the \Swallow\ project open source contribution 
(see Sect.~\ref{sect:Open_Source}).

A novel feature of this energy measurement is that the measurement data can be
offloaded to the \Swallow\ slice itself. In this way, it is possible to create a
program that can measure its own power consumption and adapt to the results.
Alternatively, the results can be streamed out of the system
using the ethernet interface. 

Due to the separate measurement points, \Swallow\ can also monitor the balance of
energy consumed by the processing cores and the external communication
channels. In a system such as ours, there are non-negligible capacitances on
the communication links, both on board and on the flexible cables. Swallow
presents a high data rate on these links and so one would expect a measurable
amount of power to be dissipated on the links. We present the 
energy per bit for each link in Tab.~\ref{tab:Link_Energies}.

The links are quite energy efficient, using approximately 100pJ/bit
for package-to-package transmission at a data rate of 125Mbit/s.
The low value can
be attributed to the sparse encoding on the link, which requires only four wire transitions to signal a byte of data. Therefore, the worst case energy usage in communication is one half that for a naïve serial or parallel link.
Once transmissions go off-board via long flexible cables, the
capacitance of those cables becomes the dominating factor for energy,
and the energy cost per bit rises by approximately $50\times$.
Shortening the cables would reduce the cost proportionately, so a more
optimised design would do this.

\begin{table}
    \centering
    \begin{tabular}{|L{1.5cm}|c|c|c|}
\hline
\textbf{Link type} & \textbf{Data Rate} & \textbf{Max Link Power} &
\textbf{Energy per bit} \\
\hline
On-die             & 500Mbit/s &  1.4mW & 1.63pJ/bit \\ \hline
On-board, vertical & 125Mbit/s & 13.3mW & 106pJ/bit \\ \hline
On-board, horizontal & 125Mbit/s & 12.6mW & 101pJ/bit \\ \hline
Off-board, 30cm FFC & 125Mbit/s  & 680mW  & 5440pJ/bit \\
\hline
\end{tabular}
\caption{Per-bit energies for three types of \Swallow\ link}
\label{tab:Link_Energies}
\end{table}

\section[Energy proportionality]{Energy proportionality}
In order for a many-core system to be scalable, its energy consumption must be
both scalable and proportional to the computation it is undertaking. Modern
high-performance computing centres consume vast quantities of energy, and
energy density issues is the most important throttler of continued integration
of more and more compute power into a fixed space. Scalable computing
systems must therefore be both energy efficient and consume energy proportionally to the compute being undertaken.
\Swallow\ is both energy efficient and energy proportional, so is scalable.

\subsection{Swallow is energy efficient}
\label{sect:core_power}
\Swallow\ is energy efficient
since the processor and memory architecture is targeted for the embedded
application space, where energy is a primary design goal. A single processor in Swallow consumes a maximum of 193mW, when active, leading to 3.1W/slice. Losses in the on-board power supplies and other support logic increase the overall power consumption to $\sim 4.5$W/slice (equivalent to 260mW/core, so a complete 480 core, 30 slice system consumes only 134W.

\begin{figure}
\includegraphics[width=\columnwidth]{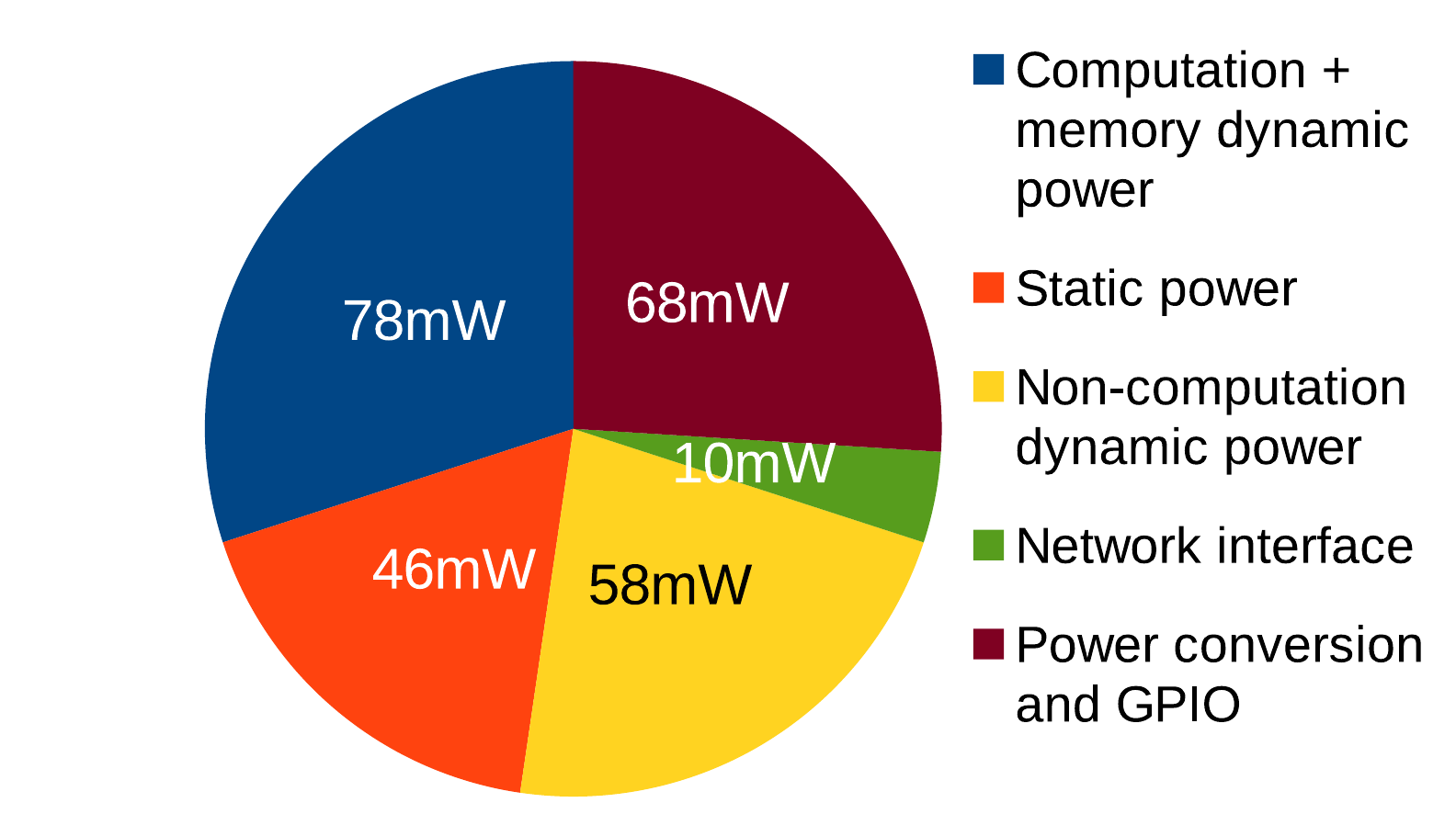}
\caption{Power consumption breakdown per Swallow node}
\label{dia:Node_Power}
\end{figure}

Overall, we see that approximately 26\% of power is used in power supply conversion, support logic and I/O,  30\% is consumed in performing computation, 40\% is wasted in non-computational static and dynamic power, and the
remaining 4\% is used for network interfacing (Fig.~\ref{dia:Node_Power}).

\subsection{Swallow is energy proportional}
\label{sect:energy_proportional}
\begin{figure}
\includegraphics[width=\columnwidth]{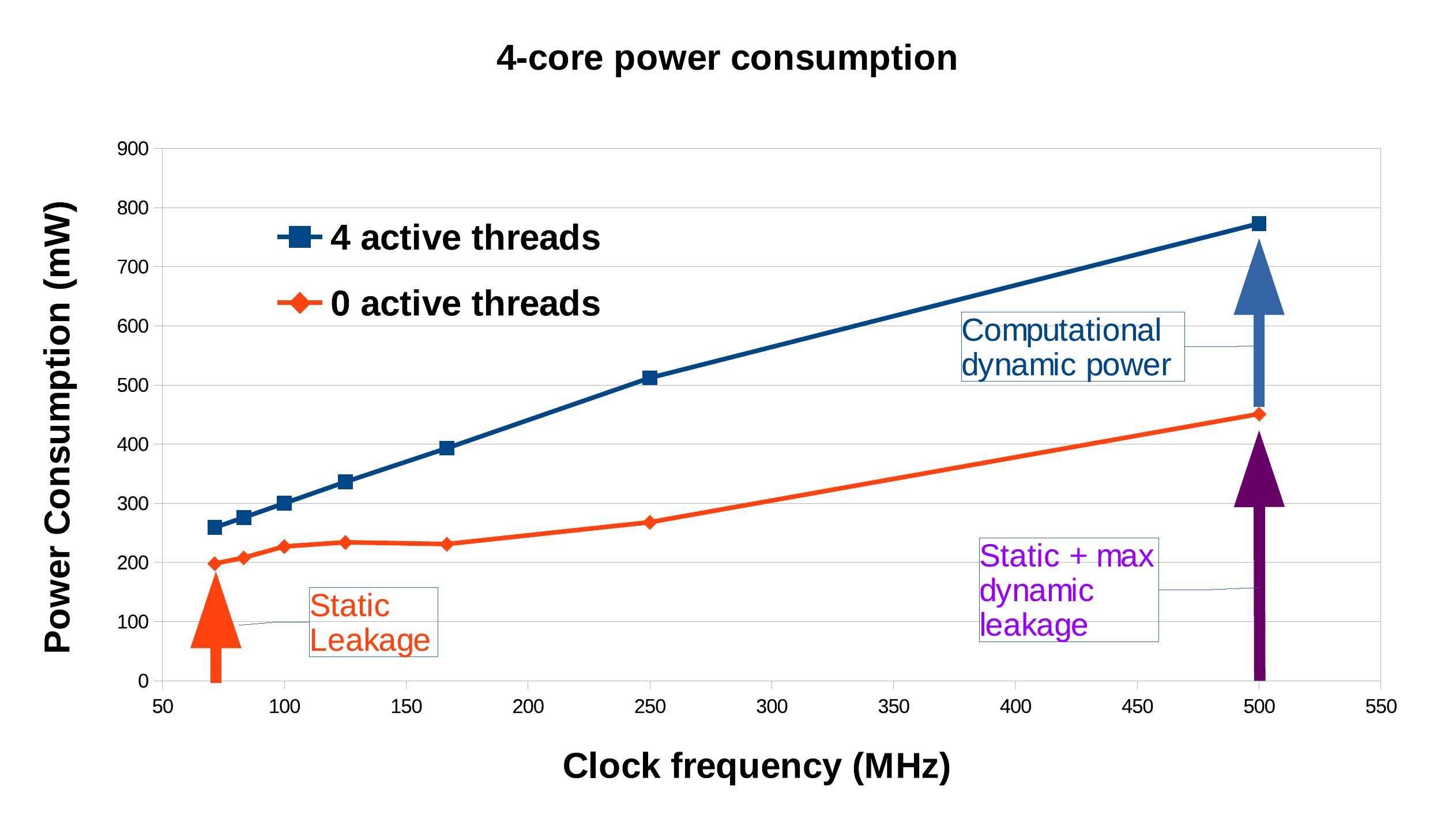}
\caption{Power consumption with dynamic frequency scaling (4 processors)}
\label{dia:Swallow_DFS_both}
\end{figure}

\Swallow\ is energy proportional since it supports dynamic frequency scaling, based on run-time load factors.
Fig.~\ref{dia:Swallow_DFS_both}
shows how power consumption of a stripe of four processors scales with the clock frequency that is set on the devices.

The power consumed per core goes from 193mW at 500MHz to 65mW at 71MHz when
cores are fully loaded with work (blue squares), and 113mW at 500MHz to 50mW at 71MHz when all cores and threads are idle (red diamonds). The characteristics are linear, giving a directly energy proportional response to clock speed $f$,
\begin{equation}
\mathrm{Power\ consumption\ per\ core} = (46 + 0.30f)\ \textrm{mW}
\label{eqn:power_vs_f}
\end{equation}
Thus, we see that the static power dissipation is 46mW/core and the dynamic
dissipation is 0.30mW/MHz.

\begin{figure}
\includegraphics[width=\columnwidth]{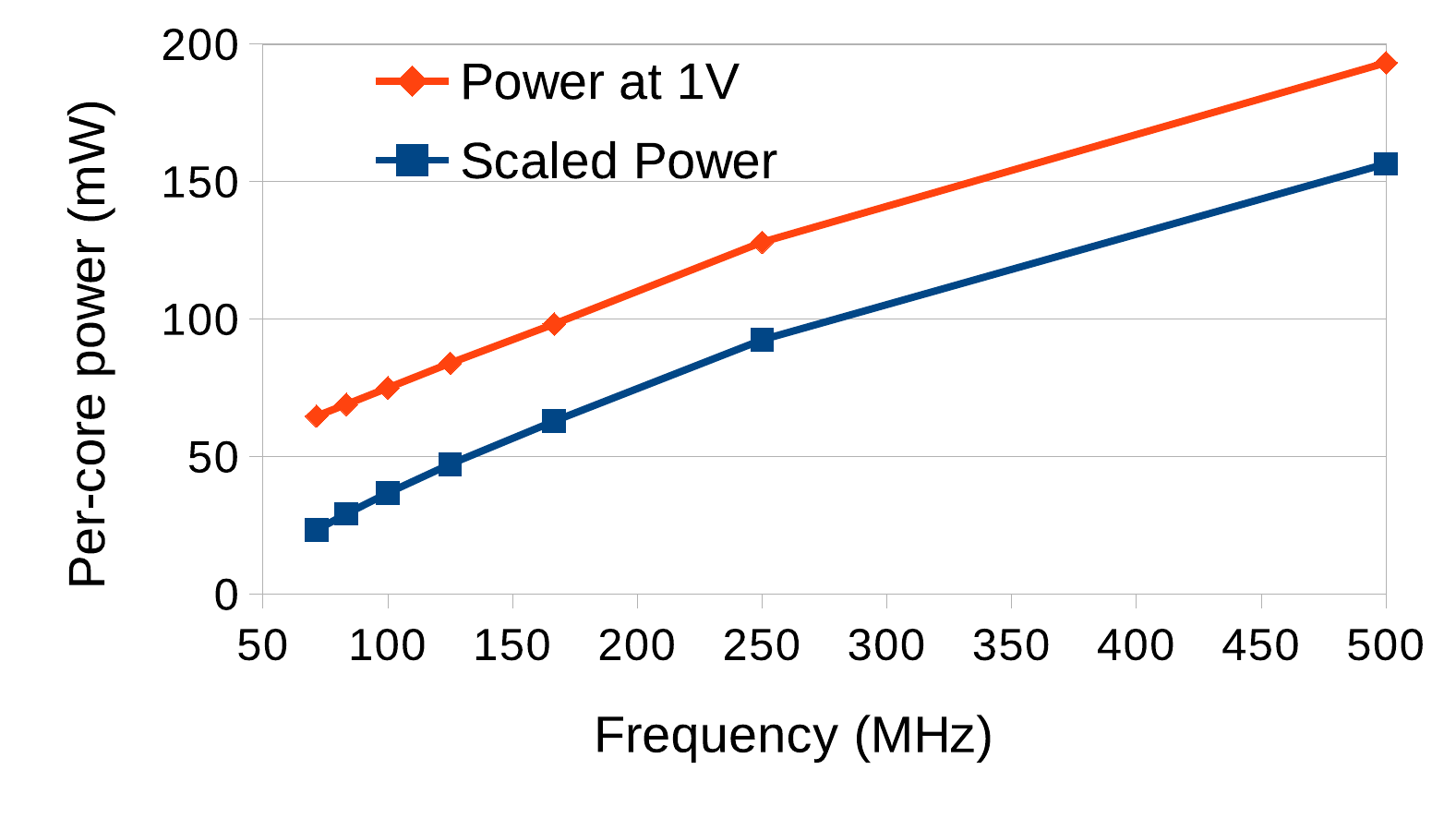}
\caption{Impact of voltage and frequency scaling on power consumption (4 active thread load)}
\label{dia:DVFS_savings}
\end{figure}

Although the current version of Swallow does not support voltage scaling, newer devices using the XS-1 ISA do support full DVFS.
The additional power savings from voltage scaling on top of frequency scaling
can be reliably calculated from knowing the power formula $P=CV^2f$, where $C$ is the
capacitance of the switching transistors and $V$ is $V_{dd}$. We have determined the minimum allowable voltage experimentally to be
0.6V at 71MHz and 0.95V at 500MHz, and calculate the equivalent DVFS savings for Swallow in Tab.~\ref{dia:DVFS_savings}.

\section{An operating system for Swallow}
A key enabler for a large-scale computing system, such as \Swallow\ is 
to abstract away implementation details such as thread creation, mapping,
network configuration and energy optimisation and just provide a programmer with
the primitives needed to construct programs in the styles that \Swallow\ targets.

Therefore, an operating system has been created for \Swallow, which provides the 
above functionality and optimises run-time scenarios with limited programmer
intervention. The operating system is called \emph{nOS} (a nano-Operating System),
and is outlined in a separate publication~\cite{Hollis:nOS}.

In other work, we are building extensions to nOS in energy measurement
and actively adapts workloads to run-time energy consumption.
\section[Related Work]{Related Work}

There are few embedded systems made at the same scale as \Swallow, with even
fewer being designed for general purpose computation.

The Tilera Tile~\cite{Bell2008} comes the closest to matching \Swallow's goals and form. The Tile 64 system comprises a 64-core system with a series of overlaid
networks to provide low latency and high throughput between cores in a software
configurable way. The effect is to provide a very agreeable $E/C$ of 2.4 with 64 cores,
and general purpose computation is supported as well as sophisticated network
traffic manipulation. The system is highly optimised for streaming traffic,
but relies on adding additional networks to improve network performance in larger
systems. This is clearly scalable in the long-run, so provides a limit for the
architecture's growth.

The Centip3De system~\cite{Fick2013} aims to use 3-D stacked dies to 
implement a 64-core system based on the ARM Cortex-M3 processor. Whilst its
scale is within an order of magnitude of \Swallow, it is clear that the 
system, which relies on a series of crossbars and coherent DRAM storage,
is not as fundamentally scalable as needed to build truly large systems.
Further, the design choices leave it with an undesirable $E/C$ ratio of 55.

In Tab.~\ref{tab:e_over_cs}, we see that the SpiNNaker system~\cite{Furber2013}
is the best provisioned system in the large scale. SpiNNaker, like Centip3De is based
on ARM cores, connecting up to one million ARM9 parts via a highly-connected network.
However, the system is targeted at solving a single problem, making it very difficult
to overlay general computation tasks, and also making it hard to draw parallels
with \Swallow\ and the other systems mentioned above.

Whilst not a modern system, the Transputer architecture~\cite{Homewood1987}
bears strong resemblence to the XMOS XS-1 architecture used by \Swallow, and
was also designed to be interconnected and scale to thousands of processing nodes
via interconnection of 100Mbit/s IEEE1355 standard links. The T400 family devices
were scaled to a 42-core machine to produce B0042 boards capable of highly parallel
computation.

\subsection{Comparison of $E/C$ ratio of contemporary many-core systems.}
We now compare \Swallow's communication-to-computation ratio compares to that of
other contemporary many-core systems. The key data is displayed in Tab.~\ref{tab:e_over_cs}.

We see that there is a wide range of $e/c$ and $E/C$ values, ranging from 0.075--55.
Notably, all systems display a minimum 14-fold drop in communication capacity when measured globally, as opposed to locally. The SpiNNaker system fares best, but is not general purpose; in this category, Swallow and Tile have the lowest reductions
in communication capacity: 16 and 32 times, respectively. 

These data illustrate the difficulties in building scalable networks of a large size, and
emphasise the importance of designing carefully for global throughput, if the application domain
requires this.

\begin{table}
    \centering
    \setlength\tabcolsep{1mm}
    \begin{tabular}{|l|R{1.8cm}|R{1.4cm}|R{1.2cm}|*{2}{R{0.65cm}|}}
\hline
\textbf{System} & \textbf{Processor source potential (bps)} & \textbf{Local sink
capacity (bps)} & \textbf{Router capacity (bps)} & $\mathbf{e}/\mathbf{c}$ &
$\mathbf{E}/\mathbf{C}$ \\
\hline
\Swallow  & 4~G    & 2~G   &  4.5~G  & 2   & 8--32  \\
SpiNNaker~\cite{Furber2013} & 6.4~M & 240~M & 4~G & 0.03 & 0.42 \\    
%
Centip3De~\cite{Fick2013} & 246~G &  ---      & 4.46~G     &    ---     & 55 \\
%
Tile~\cite{Bell2008}  & 96~G  & 1.28~T  & 2.56~T   & 0.075  & 2.4 \\
%
Epiphany~\cite{Adapteva} & 19.2~G & 2~G & 51~G  & 0.10 & 6.02 \\
%
\hline
\end{tabular}
\caption{Communication to computation ratios for contemporary many-core systems}
\label{tab:e_over_cs}
\end{table}

\subsection{Energy comparison to contemporary many-core systems}
In Tab.~\ref{tab:power_comparison}, we compare the power consumption of
\Swallow\ with a variety of contemporary many-core systems.
Its power per core is in the middle of the range, which can be
explained by the fact its operating frequency and process node
is also in the middle of the group.

The SpiNNaker device has similar characteristics per node~\cite{Furber2013}. It is designed with
lower performance ARM9 cores in 130nm, each with 64kB of data and 32kB of
instruction memory and a single DRAM per slice. It consumes 87mW of power,
averaged across its 1,036,800 processors. It is much more densely integrated
than Swallow, with 17 cores per device and is application specific. If Swallow
had these economies of scale and improved clocking, an equivalent level of
power efficiency is very possible to obtain.

The Centip3De system exploits near-threshold computing in 130nm, small ARM
Cortex-M3 cores and consumes 203---1851mW/core, depending on its configuration~\cite{Fick2013}.
Centip3De's high power cost is mainly due its cache-centric design,
which \Swallow\ has abandoned.

Tilera's 64 core device~\cite{Agarwal2007} consumes 300mW/core (19.2W/device). Adapteva's
specialised floating point Epiphany architecture is claimed to require
$<2$W for a 28nm 64-core device (31mW/core), but this data is not
currently verified.


\begin{table*}
    \centering
    \begin{tabular}{|L{1.3cm}|L{2cm}|*{3}{R{1.5cm}|}R{2.1cm}|*{2}{R{1.7cm}|}}
\hline
{\bfseries System} &
{\bfseries ISA} &
{\bfseries Cores / device} &
{\bfseries Cores / system} &
{\bfseries Technology node} &
{\bfseries Power / core} &
{\bfseries Operating speed} &
{\bfseries {\textmu}W/MHz}\\\hline
Swallow &
XMOS-XS1 &
2 &
16--480 &
65nm &
193mW [\S~\ref{sect:core_power}]&
500MHz &
300 [Eqn.\ref{eqn:power_vs_f}]\\\hline
SpiNNaker &
ARM9 &
17 &
1,036,800 &
130nm &
87mW &
200MHz &
435\\\hline
Centip3De &
ARM Cortex-M3 &
64 &
64 &
130nm &
203--1851mW &
20--80MHz &
2540--2300\\\hline
Tilera &
Tile &
64 &
64--480 &
130nm &
300mW &
1000MHz &
300\\\hline
Adapteva Epiphany &
Epiphany &
64 &
64 &
28nm &
31mW &
800MHz &
38.8\\\hline
\end{tabular}
\caption{Comparison of per-core power of contemporary many-core systems}
\label{tab:power_comparison}
\end{table*}

\section[Applications for Swallow]{Applications for Swallow}
Swallow is built as a general purpose compute system. However, its scale means
that only a selection of applications will be able to exploit all of the
parallelism that it exposes.
We have implemented algorithms such as image recognition, path searching and
network simulation, all of which map well to either the farmer-worker of
pipelined paradigms.

We do not have space to describe them all, but in this section
we introduce two case studies run on Swallow. We describe their behaviour and
and show that they scale well.

\subsection{Case study I: neural network simulation}
We created a simulation of a neural network using the Izhikevich
spiking neuron model, as adopted by Furber et al.~\cite{Furber2013}
and Moore et al.~\cite{Moore:BlueHive}.
We took the approach that a single neuron is represented by its own copy of the
state variables from the Izhikevich model, and that many neurons can be
combined onto a single core. Neurons are connected by message passing, which
simulates their spiking behaviour.
Using the event-driven
nature of the processors, neurons sleep until a stimulus arrives. This
increases the scalability of the computation by ensuring that simulation
computations need only occur for those neurons that receive new stimuli in a
simulation time-step.

A neuron's state variables are few and
take little data memory: 8 bytes of state plus 10 bytes of even buffer space. The simulation code is also compact: 1.1kBytes of shared code,
with each copy requiring 336Bytes of stack.
Thus, it would appear that the simulation is very efficiently scalable both in amalgamating many neurons onto a single core and in distributing neurons across cores. The
memory-based limit on a single core is thus 191 neurons.

In terms of memory usage, the
scaling of the number of neurons with cores can be made linear. In terms of
computational performance, it is also
linear. However, the real
limit on scaling for this
simulation is provided by
the need to communicate `spikes' between neurons.
In the Izhikevich model, a neuron may spike in response to a stimulus. Both the stimulus and spike are
represented in Swallow by the arrival and departure of messages on the network.

To provide a realistic simulation of a human brain, each neuron should be connected to at least 10\% of the neuron population (of
size $N$)~\cite{Furber2013}. Every spike that it
manufactures must be delivered as an input to each of these connected neurons.
This entails large-scale multi-cast of messages.
It is tempting to reduce this percentage, to ensure that the computation remains tractable. However, this
also renders the simulation worthless as a simulation of the human brain.
For this reason, we treat 10\% as a minimum connectivity constraint on our simulation.

Under this constraint, scaling becomes surprisingly difficult.
Whilst Swallow's network is flexible, it only implements point-to-point routing and the hardware routing
tables do not have the capacity to store the required number of entries to link
every neuron to every other neuron. So, we must use software routing, overlaid
on the point-to-point network to manufacture the multi-cast needed by spike
outputs. The interesting thing here is that the routing table then used by the
software router becomes the dominant factor in scaling the simulation, since it
must indicate which 10\% of the population of neurons are connected to each
neuron. Since each neuron has different connectivity (initialised randomly in
our simulations), each neuron must store a unique copy of this table. Under the
strategy that a connection is stored by a single bit, each neuron must keep a
table at least $N$ bits long.

Whilst this size is relatively small compared to the core memory for a single
neuron, the addition of neurons to the simulation not only results on more
neurons needed per core, but also a larger table for each neuron. This results
in a double whammy and soon one must remove neurons from cores in order to
allocate sufficient memory to each. Then, the only way to maintain scaling is
to add more processors, each simulating fewer and fewer neurons. This results
in an asymptotic limit of $P$ neurons in the simulation for large $P$.

\begin{figure}
\includegraphics[width=\columnwidth]{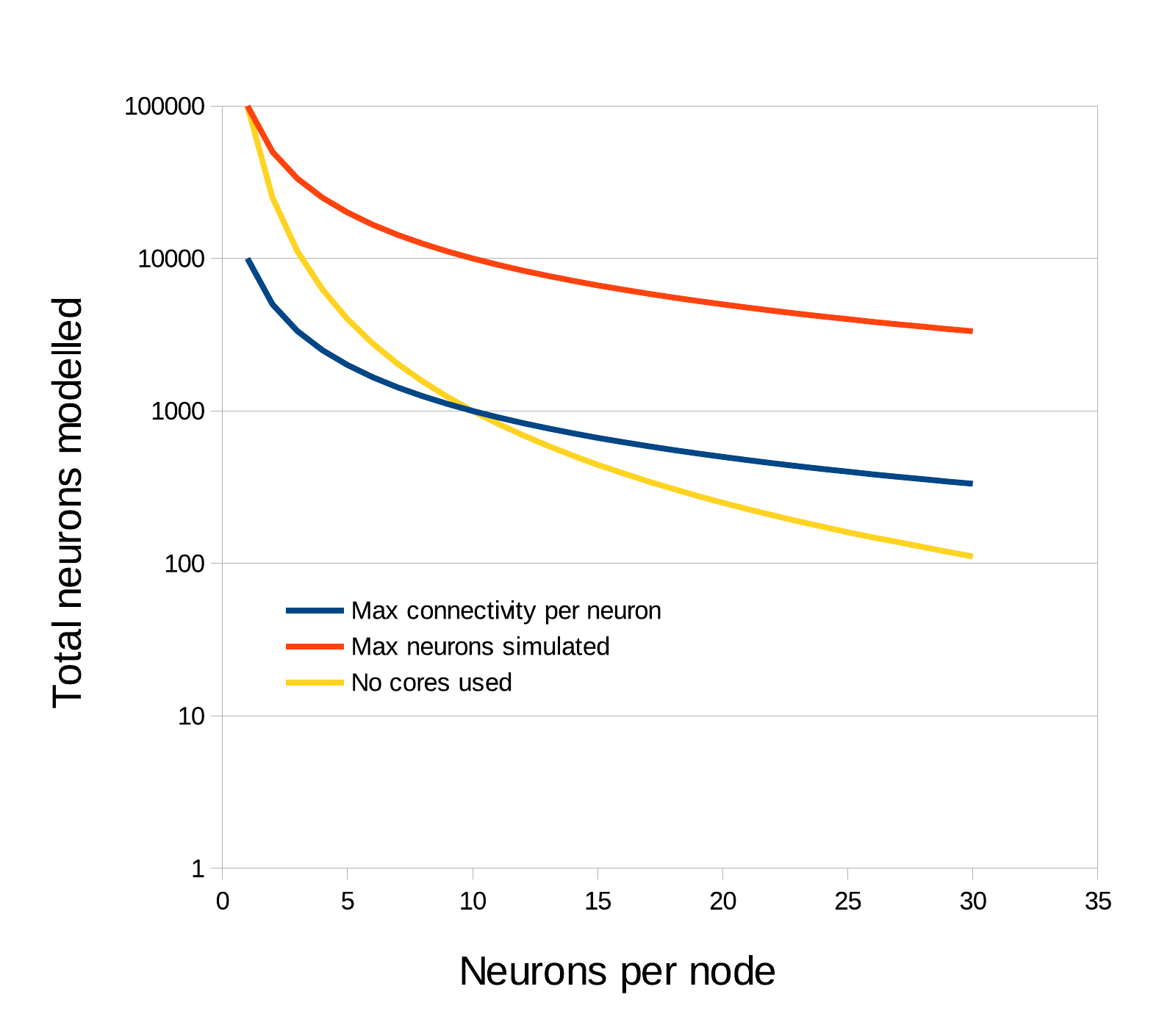}
\caption{Scalability of the Izhikevich neuron model simulation.}
\label{dia:Neuron_Scaling}
\end{figure}

The curves in Fig.~\ref{dia:Neuron_Scaling} show the effect of this scaling, under the assumption that 100,000 cores are available. 
The x-axis shows the number of neurons allocated per core. The red line then shows how the total number of neurons simulated the number
of processors needed to do so increases by $P=N^2$. For the Swallow
architecture, there is a hard memory limit of 100,000 neurons, which would
require 100,000 processors! Therefore, the actual number of neurons that we
have been able to simulate with our 480 processors is much more modest -- in
the mid thousands of neurons.

This poor scaling is well understood in the community. There have been attempts to
solve it by increasing the network connectivity (e.g. the partially connected
toroidal network of SpiNNaker), by substituting random accesses to memory for
connectivity (e.g. the large DRAM banks of BlueHive), but all approaches are
eventually constrained by the fact that connectivity cannot continually
increase at a rate of $N^2$.

Thus, Swallow has a natural limit for these kinds of problems and a large
Swallow system would be better placed running multiple simultaneous copies of
more modest neural simulations --- say 10,000 neurons, than attempting to
scale up the number of neurons in a given simulation.
Therefore, the fundamental scaling property of the simulation has
given us an insight into the most efficient realisation.

%

\subsection{Case study II: emulating shared memory}
We now show a case study that explores how \Swallow\ is powerful enough to allow the assumptions underlying the computational model to
be broken. Assume that our goal is to offer a global shared memory to our user programs. It can be done in a number of
ways, but one efficient way to is to add a single large memory store and
allocate a thread of execution as a memory  controller that responds to read and write requests to memory locations. 

The thread and channel IDs of such a controller can be passed to all instances
of nOS, which then support user calls such as \ttt{shared\_read(address)} and
\ttt{shared\_write(address)} by translating them into communication to and from
the controller.

Of course, this strategy has a single point of contention and, in the presence
of frequent transfers can also cause congestion in the network.

A more elegant strategy emulates shared memory using a collection of $n$
distributed memories of size
$m_0$\ldots$m_{n-1}$ bits to produce a total capacity of
size
$\sum_{m_0}^{m_{n-1}} m_i$ bits. The realisation
of such involves each nOS instance acting as a memory controller for a sub-section of address space. In the optimal case,
$m_0$\ldots$m_{n-1}$ are equal and the controller can be identified simply by calculating the nOS instance ID as
\ttt{address}\%$n$.

\section{Open source release}
\label{sect:Open_Source}
\Swallow\ is an open source project and will be released under the CC-BY-SA 4.0 International licenses (for design elements and documentation). 
The Swallow nOS operating system is released under GPLV3 at
\url{https://github.com/simonhollis/swallow-nos}. 
Support code and libraries for the hardware platform will be released under a
compatible license in the near future.

The latest status of all releases can be found at \url{http://www.cs.bris.ac.uk/Research/Micro/swallow.jsp}.

\section{Conclusions}
We have presented a new processing system architecture named \Swallow.
\Swallow\ is designed to be inherently scalable to thousands of cores.
This led to design decisions, such as distributed memory, a scalable network on-
and off-chip implementation, with a best case computation to communication
ratio of 8. 

Swallow is modular and our example system comprises 480 cores and network nodes.
It is highly energy efficient, using only 193mW/core with four active threads.
Swallow supports dynamic frequency scaling and this allows an energy reduction to as
little as 50mW/core when idle.

Key application paradigms targeted by \Swallow\ include farmer-worker and computational
pipelines and case studies have been shown to motivate this application space.
We have also shown how large data sets and large application code bases can be
supported even when individual processing nodes have limited storage.
A distributed operating system has also been designed to ease programmer interaction
and make run-time optimisations.

\bibliographystyle{IEEEtran}
\bibliography{Swallow}


\end{document}